%% file: isit2021.tex
\documentclass[conference]{IEEEtran}
\IEEEoverridecommandlockouts
\input{./preamble}

\usepackage{textcomp}

\begin{document}
\title{Multi-Class Unsourced Random Access\\via Coded Demixing}

\author{\dag Vamsi K. Amalladinne,
\S Allen Hao,
\S Stefano Rini,
\dag Jean-Francois Chamberland\\
\dag Department of Electrical and Computer Engineering, Texas A\&M University\\
 \S Department of Electrical and Computer Engineering, 
 National Yang Ming Chiao Tung University (NYCU)
\thanks{
This material is based upon work supported, in part, by the National Science Foundation (NSF) under Grant CCF-1619085, and by Qualcomm Technologies, Inc., through their University Relations Program.}
}

\maketitle

\begin{abstract}
Unsourced random access (URA) is a recently proposed communication paradigm attuned to machine-driven data transfers. 
%
In the original URA formulation, all the active devices share the same number of bits per packet.
The scenario where several classes of devices transmit concurrently has so far received little attention.
An initial solution to this problem takes the form of group successive interference cancellation, where codewords from a class of devices with more resources are recovered first, followed by the decoding of the remaining messages.
%
This article introduces a joint iterative decoding approach rooted in approximate message passing.
This framework has a concatenated coding structure borrowed from the single-class coded compressed sensing and admits a solution that offers performance improvement at little added computational complexity.
Our findings point to new connections between multi-class URA and  compressive demixing. 
%
%
%
%
%
%
The performance of the envisioned algorithm is validated through numerical simulations. 
\end{abstract}


\section{Introduction and Background}
\label{section:Introduction}
%
Machine-type communication (MTC) is poised to increase sharply over the coming decades~\cite{liu2018sparse,munari2020grant}.
This poses a great challenge to the foundation on which wireless infrastructures are relying,
as MTC often produces short packets and fleeting connections.
This results in a 
constantly changing collection of devices wishing to transmit bursty data.
In this context, it becomes impractical to adopt the 
enrollment-estimation-scheduling paradigm, as the cost of acquiring side information on the channel state becomes prohibitive.
%
This reality is reflected in the current interest in grant-free schemes and uncoordinated wireless access.

Another aspect of coordinated system that is hard to maintain in MTC is the implicit agreement between an access point and an active device regarding the MAC layer configuration. 
%
This consideration serves as a motivation behind the unsourced random access (URA) model introduced by Polyanskiy~\cite{polyanskiy2017perspective}.
A distinguishing feature of this approach is that active devices must share a same encoding function.
The signal sent by a device therefore depends solely on the message payload, and not on its identity.
If a device wishes to be recognized, then it must embed an identifier within its message.
Altogether, the URA model captures several fundamental aspects of MTC: this has led to 
%
%
a rapid growth of literature on the topic
~\cite{ordentlich2017low,marshakov2019polar,calderbank2020chirrup,amalladinne2020coded,fengler2019sparcs-isit,amalladinne2020AMP,facenda2020efficient,decurninge2020tensor,shyianov2020massive}.

An aspect of URA that has received little consideration is the heterogeneous scenario where various types of device are present.
In such situations, devices from distinct groups may wish to communicate messages of different lengths and they may have varying power budgets.
This reality can be accommodated by allowing group-based encoding and separating groups in signal space \cite{hao2020exploration}.
Interestingly, standard URA admits a compressed sensing (CS) interpretation where signals sent by active devices are collectively viewed as sparse indices in the space of possible messages.
Under this viewpoint, the heterogeneous URA framework becomes similar to 
compressive demixing in very large dimensional spaces \cite{mccoy2014sharp}.
In this article, we combine recent advances in coded compressed sensing~\cite{amalladinne2020coded,fengler2019sparcs-isit,amalladinne2020AMP} and established concepts in demixing~\cite{chen2001atomic,boyd2011distributed,hegde2012signal,amelunxen2014living,mccoy2014convexity} to develop a novel communication scheme for the two-class URA setting.

\section{System Model and Encoding Process}
\label{sec:System Model and Encoding Process}

In this section, we begin by describing the two-class URA architecture.
%
%
Consider the scenario in which $K_1$ active devices form group~1 and $K_2$ devices form group~2, each user wishing to send a message to the access point.
These transmissions take place over an uncoordinated multiple access channel.
Active devices are cognizant of frame boundaries and they collectively signal over $n$ channel uses (real degrees of freedom).
The signal received at the destination is of the form
\begin{equation} \label{equation:ChannelModel}
\textstyle \yv = \sum_{i=1}^{K_1} \xv_i^{(1)} + \sum_{i=1}^{K_2} \xv_i^{(2)} + \zv
\end{equation}
where signal $\xv_{\iota}^{(g)} \in \mathcal{C}_g \subset \mathbb{R}^n$ is the codeword transmitted by active device~$\iota \in [1,\ldots,K_g]$ in group~$g\in \{1,2\}$, and $\mathcal{C}_g$ is the codebook associated with the 
group $g$.
The noise component $\zv$ is composed of independent Gaussian elements, each with distribution $\mathcal{N}(0,1)$.
Adhering to the URA framework, we emphasize that all the active devices within a group share a same codebook; consequently, $\xv_{\iota}^{(g)}$ is a function of the payload of device~$\iota$, and not of its identity.

Our proposed algorithm loosely parallels coded compressed sensing (CCS).
The selection of a codeword mimics the encoding process obtained by combining the outer code of Amalladinne et al.~\cite{amalladinne2018coupled,amalladinne2020coded} and the SPARC-inspired inner encoding of Fengler et al.~\cite{fengler2019sparcs-isit,fengler2019sparcs}, albeit we must account for the two groups.
More specifically, information bits are encoded using a concatenated structure as in~\cite{amalladinne2020AMP,amalladinne2020unsourced}.
The outer code is an LDPC code over a large alphabet.
Every output symbol is turned into an index vector, which contains zeros everywhere except for a single location where it features a one.
These index vectors are concatenated into a SPARC-like vector~\cite{joseph2013fast,rush2017capacity,venkataramanan2019sparse}, which is subsequently multiplied by a sensing matrix.
Such a construction is somewhat intricate, and it can hardly be explained in detail within a short article.
We point the reader to~\cite{amalladinne2020coded} for an extensive description of the scheme.
Below, we  shall only provide 
a synopsis of the approach and, then, we emphasize the modifications that must be performed to accommodate two groups of devices.

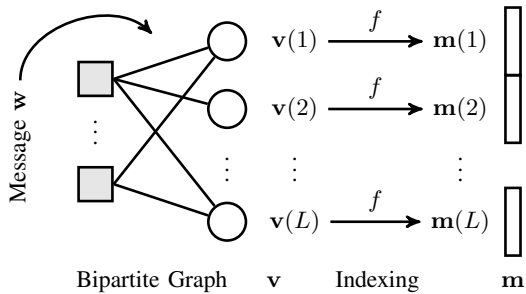
\begin{figure}[bth]
  \centering
  \input{Figures/LDPC}
  \caption{
  A depiction of the CCS encoding process in Sec. \ref{sec:System Model and Encoding Process}.
  The function $f(\cdot)$ in \eqref{equation:IndexFunction} maps strings of bits to index vectors.}
  \label{figure:MessageEncoding}
\end{figure}

Consider a sequence of information bits $\wv \in \{0, 1\}^w$.
This sequence $\wv$ is encoded using an LDPC code, 
leading to a codeword $\vv$ which is then partitioned into $L$ blocks $\vv(\ell)$ of length $v = w + p$, so that 
$\vv = {\vv(1)}  {\vv(2)} \cdots {\vv(L)}$.
This LDPC code is designed to permit efficient decoding of several codewords on a same graph using belief propagation (BP) through the application of fast Fourier transform (FFT) or similar techniques~\cite{goupil2007fft}.
As part of the next encoding step, each sub-block $\vv(\ell)$ is transformed into a message index of length $m = 2^{v/L}$.
Mathematically, we have
\begin{equation} \label{equation:IndexFunction}
\begin{split}
\mv(\ell) &= f(\vv(\ell)),
\end{split}
\end{equation}
where the function $f(\cdot)$ is a bijection between $\{ 0, 1 \}^{v/L}$ and standard basis elements of length $m$.
The SPARC-like message $\mv$ is subsequently obtained 
as the concatenation of these index vectors, that is $\mv = \mv(1) \mv(2) \cdots \mv(L)$.
In the original URA formulation, the binary vector $\mv$ is leveraged to generate transmit signal as $\Phim \mv$.
This is illustrated in Fig.~\ref{figure:MessageEncoding}.
The signal aggregate produced by all the active devices then assumes the form $\Phim \sv$, where $\sv = \sum_i \mv_i$ is the sum of the SPARC-like messages from all the active devices.

To accommodate two groups of devices, we must expand this construction.
This warrants the creation of two distinct codes $\mathcal{C}_1$ and $\mathcal{C}_2$, both of which follow an encoding architecture akin to Fig.~\ref{figure:MessageEncoding}.
Thus, the received signal becomes
\begin{equation} \label{equation:InnerChannel}
\begin{split}
\yv &= \Phim_1 \sv_1 + \Phim_2 \sv_2 + \zv .
\end{split}
\end{equation}
Signal $\sv_1$ is the sum of the SPARC-like vectors $\mv^{(1)}_i$ encoded using $\mathcal{C}_1$; and $\sv_2$ is the sum of the vectors $\mv^{(2)}_i$ encoded using $\mathcal{C}_2$.
We note that both $\sv_1$ and $\sv_2$ possess sparse structures.
This hints at a connection between \eqref{equation:InnerChannel} and
compressive demixing.

\begin{remark}
There are at least two ways to arrive at \eqref{equation:InnerChannel}.
The simplest avenue is to encode the two groups separately, with devices in group~1 leading to $\sv_1$ and devices in group~2, $\sv_2$.
The second option is to create layers as in \cite{hao2020exploration}.
Suppose that messages within group~1 are smaller than those in group~2, i.e., $w_1 < w_2$.
Then, all the devices in group~1 employ $\mathcal{C}_1$.
On the other hand, the devices in group~2 encode their first $w_1$ bits with $\mathcal{C}_1$ and, after adding linking parities, they encode the remaining $w_2 - w_1$ bits using $\mathcal{C}_2$.
These distinct approaches lead to essentially equivalent mathematical formulations, up to parameterization.
Therefore, we discuss the simpler form with the understanding that findings extend to the two alternatives.
\end{remark}

\section{Demixing and Decoding Algorithm}

Having established the encoding for our multi-class URA problem, we now turn our attention to the decoding process.
Conceptually, the receiver seeks to recover $\sv_1$ and $\sv_2$; and it must disentangle the information messages embedded in each aggregate.
The first task can be viewed as support recovery through 
compressive demixing, and the second objective corresponds to message disambiguation.
As in~\cite{amalladinne2020AMP}, we propose a decoding framework that performs these two undertakings concurrently.
The building blocks for our algorithm include BP on factor graphs~\cite{kschischang2001factor} and approximate message passing (AMP) applied to non-separable functions~\cite{berthier2020state}.

\subsection{Belief Propagation (BP)}

The application of 
BP to bipartite LDPC graphs is well understood~\cite{kschischang2001factor} and we shall only briefly review it here. 
It can be summarized as follows: $\lambdav_{\ell}$ denotes the trivial factor associated with a local observation; $\muv_{s \to a}$ represents the message from variable~$s$ to factor~$a$; and, $\muv_{a \to s}$ is the message from factor~$a$ to node~$s$.
Under suitable conditions, 
BP efficiently computes marginal probabilities at every variable node~$s$.
\begin{figure}[tbh]
  \centering
  \input{Figures/BP}
  \caption{
  Our scheme employs factor graphs over large alphabets, one per group, to partially recover several codewords at once.}
  \label{figure:BeliefPropagation}
\end{figure}
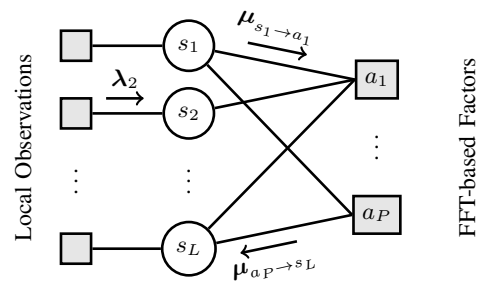

Still, there is an important distinction between standard decoding on factor graph and its use in URA settings.
The decoder seeks to recover multiple messages on a same graph simultaneously.
This distinguishing aspect and its implications are carefully described in~\cite{amalladinne2020unsourced}.
Note that it forces the LDPC code to be over a very large alphabet.
Also, it changes the way $\{ \lambdav_{\ell} \}$ are initialized, a consideration we delayed until the next section.
Nevertheless, we can interpret vector $\lambdav_{\ell}$ for one group as a collection of local estimates.
Entry $\lambdav_{\ell}(k)$ is an estimate for the event that a designated device has produced basis element $\ev_k$ within section~$\ell$, i.e., a proxy for $\Pr (\mv_i(\ell) = \ev_k)$.

Once trivial factor nodes are initialized, the message passing rules are essentially standard.
A message going from check node~$a_p$ to variable node~$s$ is given by
\begin{equation} \label{equation:BP-Check2Variable}
\muv_{a_p \to s} (k)
= \sum_{\kv_{a_p}: k_p = k} \mathcal{G}_{a_p} \left( \kv_{a_p} \right)
\prod_{s_j \in N(a_p) \setminus s} \muv_{s_j \to a_p} (k_j),
\end{equation}
where $N(a_p)$ is the graph neighborhood of $a_p$
and indicator function $\mathcal{G}_{a_p} (\cdot)$ assesses the local consistency of the index sub-vector $\kv_{a_p} = ( k_s : s \in N(a_p) )$.
Similarly, a message passed from variable node $s_{\ell}$ to check node $a$ can be expressed as
\begin{equation} \label{equation:BP-Variable2Check}
\muv_{s_{\ell} \rightarrow a} (k)
\propto \lambdav_{\ell} (k) \prod_{a_p \in N(s_{\ell}) \setminus a} \muv_{a_p \to s_{\ell}} (k),
\end{equation}
where the `$\propto$' symbol indicates that the measure should be normalized.

At any point in the iterative process, the belief vector on section~$\ell$ based on extrinsic information is proportional to the product of the messages from adjoining check factors, i.e.
\begin{equation} \label{equation:EquivalentPriors}
\muv_{s_{\ell}}(k) = \prod_{a \in N(s_{\ell})} \muv_{a \to s_{\ell}} (k) .
\end{equation}
The estimated marginal distribution of a specific device having transmitted index~$k$ at variable node~$s_{\ell}$ is proportional to the product of the current messages from all adjoining factors, including its intrinsic information, that is 
\begin{equation} \label{equation:BP-TildeM}
\begin{split}
p_{s_{\ell}} (k) &\propto \lambdav_{\ell} (k) \prod_{a \in N(s_{\ell})} \muv_{a \to s_{\ell}} (k)
= \lambdav_{\ell} (k) \muv_{s_{\ell}} (k) .
\end{split}
\end{equation}
A normalized version of $\lambdav_{\ell} (k) \muv_{s_{\ell}} (k)$ can be viewed as an estimate for probability $\Pr \left( \mv_i(\ell) = \ev_k \right)$ (within a group), where $i$ is fixed.

Again, in the context of multi-class URA and \eqref{equation:InnerChannel}, there are actually two factor graphs: one associated with $\mathcal{C}_1$ and the other, with $\mathcal{C}_2$.
BP
is similar in these two cases, which justifies our unified treatment.
Below, we draw a distinction between the two codes wherever appropriate.
It is also worth noting that, during joint decoding, we utilize 
BP in an unusual way with frequent reinitialization and only a few steps of 
BP per round.
The motivation behind this strategy is explained below.

\subsection{Approximate Message Passing (AMP)}

In this section, we explain the component of the decoding algorithm associated with the inner code.
The mathematical structure in \eqref{equation:InnerChannel} hints at several demixing algorithms found in the literature to address similar tasks \cite{boyd2011distributed,hegde2012signal,amelunxen2014living,mccoy2014convexity}.
Yet, as mentioned in the introduction, a distinguishing feature of our setting lies in the very large dimensions of the raw index vectors.
Much like in the case of CCS, this precludes the direct application of existing demixing methods.
As such, we strive to create schemes that harness lessons from the past while also admitting a computationally tractable implementation.
Specifically, we wish to leverage insights from demixing and transpose them into an AMP framework, which has recently been adapted to the standard URA formulation~\cite{fengler2019sparcs-isit,amalladinne2020AMP}.
The appeal of AMP-based solutions stems, partly, from their low-complexity and mathematical tractability~\cite{donoho2009message,bayati2011dynamics}.

To conform to the structure of typical AMP formulations, we rewrite the received signal in \eqref{equation:InnerChannel} as
\begin{equation} \label{equation:InnerChannel2}
\begin{split}
\yv 
= \Am \begin{bmatrix} d_1 \sv_1 \\ d_2 \sv_2 \end{bmatrix} + \zv
= \begin{bmatrix} \Am_1 & \Am_2 \end{bmatrix}
\begin{bmatrix} d_1 \sv_1 \\ d_2 \sv_2 \end{bmatrix} + \zv .
\end{split}
\end{equation}
where $d_g \Am_g = \Phim_g$.
In this abstraction, matrix $\Am$ has independent, zero-mean Gaussian entries.
The variance of these entries is selected as to normalize the 2-norm of every column, $\mathbb{E} \left\| \Am_{:,\iota} \right\|^2 = 1$.
Constant $d_g$ is a non-negative amplitude; it accounts for the transmit power employed by devices in each group.
We emphasize that \eqref{equation:InnerChannel2}, on top of having a standard AMP format, matches the canonical compressive demixing equation found in~\cite{amelunxen2014living}, albeit with an additional noise term.
In its current form, this representation invites a composite iterative AMP decoder that alternates between \eqref{equation:AMP-Residual} and \eqref{equation:AMP-Denoising}:
\begin{gather}
\zv^{(t)} = \yv - \Am \begin{bmatrix} d_1 \sv_1^{(t)} \\ d_2 \sv_2^{(t)} \end{bmatrix} + \frac{\zv^{(t-1)}}{n} \operatorname{div} \etav^{(t-1)} \left( \rv^{(t-1)} \right) \label{equation:AMP-Residual} \\
\begin{bmatrix} \sv_1^{(t+1)} \\ \sv_2^{(t+1)} \end{bmatrix}
= \etav^{(t)} \left( \rv^{(t)} \right)
= \begin{bmatrix} \etav_1^{(t)} \left( \rv_1^{(t)} \right) \\ \etav_2^{(t)} \left( \rv_2^{(t)} \right) \end{bmatrix}, \label{equation:AMP-Denoising}
\end{gather}
where the \emph{effective observation} $\rv^{(t)}$ is given by
\begin{equation} \label{equation:Effective-Observation}
\rv^{(t)} = \begin{bmatrix} \rv_1^{(t)} \\ \rv_2^{(t)} \end{bmatrix}
= \Am^{\mathrm{T}} \zv^{(t)} + \begin{bmatrix} d_1 \sv_1^{(t)} \\ d_2 \sv_2^{(t)} \end{bmatrix} .
\end{equation}
The iterative process is initiated with $\sv^{(0)} = \zerov$ and $\zv^{(0)} = \yv$.
Equation \eqref{equation:AMP-Residual} is the computation of the \emph{residual} augmented by the characteristic AMP Onsager term~\cite{bayati2011dynamics}.
The second equation is a state update, which is performed through denoising using function $\etav^{(t)} (\cdot)$.
The remainder of this section is devoted to defining and justifying the denoising functions.

An astounding property of AMP is that, under certain conditions, the effective observation is asymptotically distributed as the true state plus Gaussian noise.
For the problem at hand, in the limit where the system gets larger, $\rv^{(t)}$ becomes approximately distributed as
\begin{equation}
\begin{bmatrix} d_1 \sv_1 \\ d_2 \sv_2 \end{bmatrix} + \tau_t \zetav_t,
\end{equation}
where the entries in $\zetav_t$ are independent $\mathcal{N}(0,1)$ random variables and $\tau_t$ is a deterministic quantity.
This phenomenon underlies the rationale behind our denoising functions.

The first ingredient of our denoiser is the posterior mean estimate (PME) of Fengler et al.~\cite{fengler2019sparcs}, a low-complexity estimator that encourages sparsity in $\sv^{(t)}$.
Their work advocates an element-wise estimate of the form
\begin{equation} \label{equation:OriginalPME}
\hat{s}_{g} \left( q, r, \tau \right)
= \frac{q \exp \left( - \frac{ \left( r - d_{g} \right)^2}{2 \tau^2} \right)}
{ q \exp \left( - \frac{ \left( r -  d_{g} \right)^2}{2 \tau^2} \right)
+ (1-q) \exp \left( -\frac{r^2}{2 \tau^2} \right)}
\end{equation}
where $q$ is the prior probability of entry $s$ being equal to one.
As its name suggests, $\hat{s}_{g} \left( q, r, \tau \right)$ is equal to the mean of $s$ under the aforementioned Gaussian approximation $r \sim d_g s + \tau \zeta$.
This approach performs well for the single-class URA problem, especially in view of its low complexity.

The PME denoiser found in~\cite{fengler2019sparcs} employs uninformative priors based on the known sparsity level of $\sv$.
The approach was subsequently improved after realizing that priors can be enhanced by performing one round of
BP on the factor graph of the outer LDPC code~\cite{amalladinne2020unsourced}.
Specifically, in this latter case, $\lambdav_{\ell}$ is initialized based on the PME of the corresponding entries in $\sv$.
This vector estimate is then refined using \eqref{equation:BP-Check2Variable} and \eqref{equation:BP-Variable2Check}.
Finally, the distribution afforded by \eqref{equation:EquivalentPriors} is then utilized as priors for a final round of PME computation, leading to the output of the denoiser.
Details are available in~\cite{amalladinne2020unsourced} and, as such, we do not reproduce them in this article.
Conceptually, the enhanced version provides a means to embed the local structure of the outer code in the AMP iterate.
Accordingly, it can improve performance significantly while adding little to the computational complexity of the CCS approach.

The dynamic PME denoiser can be adapted to the current context, with its multiple classes.
We propose the following candidate implementation.
As in \eqref{equation:AMP-Denoising}, we use two instances of the denoiser, one for each group,
\begin{equation} \label{equation:PME-Denoiser}
\etav_g^{(t)}(\rv_g) \quad g \in \{ 1, 2 \} .
\end{equation}
These functions are the dynamic PME denoiser defined by
\begin{equation} \label{equation:PME-decomposition}
\etav_g^{(t)}(\rv_g) = \hat{\sv}_{g,1}(\rv_g, \tau_t) \cdots \hat{\sv}_{g,L_g}(\rv_g, \tau_t) .
\end{equation}
The number of sections in \eqref{equation:PME-decomposition} matches the number of symbols produced by the LDPC code in $\mathcal{C}_g$ (see Fig.~\ref{figure:MessageEncoding}).
Furthermore, individual entries are given by
\begin{equation*} 
\hat{\sv}_{g,\ell}(\rv_g, \tau_t)
= \left( \hat{s}_{g} \left( \qv(\ell, k), \rv(\ell, k), \tau_t \right) : k \in 0, \ldots, m_{g} - 1 \right),
\end{equation*}
where $\{ \qv(\ell, k) \}$ are the 
BP extrinsic estimate of \eqref{equation:EquivalentPriors} employed as prior probabilities in the PME.

An important subtlety in extending previous results to the current, more elaborate scenario is that $\tau_t$ is shared by the two parallel denoisers $\etav_1^{(t)}$ and $\etav_2^{(t)}$.
It must therefore be computed (or evaluated) through the joint state evolution, leading to
\begin{equation*}
\begin{split}
\tau_{t+1}^2
= \sigma^2 + \lim_{n \rightarrow \infty} \frac{1}{n} \mathbb{E} \left\| \Dm \etav^{(t)}  \left( \begin{bmatrix} d_1 \sv_1 \\ d_2 \sv_2 \end{bmatrix} + \tau_{t} \zetav_{t} \right) - \begin{bmatrix} d_1 \sv_1 \\ d_2 \sv_2 \end{bmatrix} \right\|^2,
\end{split}
\end{equation*}
where $\Dm = \operatorname{diag} (d_1 \IDm, d_2 \IDm)$ accounts for the signal amplitudes and $\sigma^2$ is the noise variance in \eqref{equation:ChannelModel}.
In practice, $\tau_t$ is often approximated by $\tau_t^2 \approx \| \zv^{(t)} \|^2/n$.
A second nuance in AMP for multi-class URA is the computation of the Onsager correction.
To get the proper form, we note that
\begin{equation} \label{equation:PME-Onsager}
\operatorname{div} \Dm \etav^{(t)} (\rv)
= d_1 \operatorname{div} \etav_1^{(t)}(\rv_1) + d_2 \operatorname{div} \etav_2^{(t)}(\rv_2) .
\end{equation}
Adapting results from \cite[Prop.~8]{amalladinne2020coded}, we then obtain
\begin{equation} \label{equation:PME-OnsagerCorrection}
\operatorname{div} \Dm \etav^{(t)}(\rv)
= \frac{1}{\tau_t^2} \left( \big\| \Dm^2 \etav^{(t)}(\rv) \big\|_1 - \big\| \Dm \etav^{(t)}(\rv) \big\|^2 \right) .
\end{equation}
This compact form takes advantage of the identities
\begin{align*}
\big\|\Dm \etav^{(t)}(\rv) \big\|^2 &= \big\| d_1 \etav_1^{(t)}(\rv_1) \big\|^2 + \big\| d_2 \etav_2^{(t)}(\rv_2) \big\|^2 \\
\big\|\Dm^2 \etav^{(t)}(\rv) \big\|_1 &= \big\| d_1^2 \etav_1^{(t)}(\rv_1) \big\|_1 + \big\| d_2^2 \etav_2^{(t)}(\rv_2) \big\|_1 .
\end{align*}
These results crucially assume that the number of 
BP iterations on each LDPC factor graph is strictly less than its shortest cycle.
The computation of the Onsager correction would become more involved otherwise.
In practice, this is not an issue because every AMP composite step entails the initiation of the local observations followed by a single round of 
BP on each graph.
Expression \eqref{equation:PME-OnsagerCorrection} can be substituted as the Onsager term in \eqref{equation:AMP-Residual}, thus completing the description of the AMP algorithm for the two-class setting.

\subsection{Disambiguation}

Once the AMP algorithm has converged, the messages embedded within $\hat{\sv}_1$ and $\hat{\sv}_2$ still need to be disentangled.
The presence of the LDPC-based message passing within the denoising functions encourages local consistency, but may fall short of decoding.
To address this issue, it suffices to run a version of the outer decoding algorithm first introduced in~\cite{amalladinne2020coded}.
Starting from root sections, the algorithm keeps track of all possible paths.
A message is added to the output list when it fulfills all the factors of the outer code.
This step is somewhat standard, as it is done separately for $\mathcal{C}_1$ and $\mathcal{C}_2$.
Additional details can be found in \cite{amalladinne2020coded,amalladinne2020AMP}.

\section{Practical Considerations and Performance}

As mentioned above, AMP is a powerful conceptual framework for the design and analysis of efficient algorithms.
Still, in practice, it can be beneficial to deviate slightly from theoretical underpinnings, especially for exceedingly large spaces.
Certain decisions can greatly reduce computational complexity.

\subsection{Implementation Details}

One modification commonly found in the CCS literature is to create $\Am$ by sampling rows of a large Hadamard matrix (except the row composed of all ones), rather than generate it randomly.
Empirical evidence suggests that this does 
not significantly affect performance, while 
circumventing the need to store a large matrix.
Furthermore, the matrix multiplication steps in \eqref{equation:AMP-Residual} and \eqref{equation:Effective-Observation} can then be performed using the fast Walsh-Hadamard transform.

For the problem at hand, we could create $\Am$ in a similar manner.
Yet, inspecting \eqref{equation:Effective-Observation} and \eqref{equation:PME-Denoiser}, we gather that $\rv_1^{(t)}$ and $\rv_2^{(t)}$ are employed separately within the denoiser.
It therefore suffices to produce these sub-vectors independently.
Thus, $\Am_1$ and $\Am_2$ can be generated in a disjoint fashion, possibly each through the sampling of a matrix amenable to fast transform techniques.
This approach retains the structural properties of the problem, and lowers the computational load slightly.
Nevertheless, it is important that $\Am_1$ and $\Am_2$ maintain good incoherence properties.
Natural candidates for this step include standard Hadamard matrices, but also extend to matrices constructed using second-order Reed-Muller functions~\cite{howard2008fast}.
These latter matrices feature good correlation properties and they have appeared in the context of single-class CCS in the past~\cite{calderbank2020chirrup}, although their use therein is quite different.

\subsection{Numerical Simulations}

We compare the performance of the proposed multi-class approach against two baselines for multi-class CCS.
The first implementation executes (group) \emph{treating interference as noise} (TIN).
That is, a single-class recovery algorithm for $\mathcal{C}_1$ is applied to $\yv$; similar steps are followed for $\mathcal{C}_2$, again using $\yv$ as the input.
This scheme would arise naturally if the two device groups were assigned to distinct access points that are not cooperating, but nevertheless share a same spectral band.
The second multi-class approach is group \emph{successive interference cancellation} (SIC).
Therein, a single-class recovery algorithm for $\mathcal{C}_1$ is applied to $\yv$, which yields estimate $\hat{\sv}_1$.
Interference from group~1 is then subtracted from observation vector $\yv$, leading to updated observation $\yv_2 = \yv - \Am_1 \hat{s}_1$.
A single-class algorithm for $\mathcal{C}_2$ is subsequently used on the refined observation $\yv_2$.
The third option is the novel multi-class CCS-AMP algorithm introduced in this article.
It can be categorized as joint decoding (JD).

The system parameters for these three candidate schemes are identical.
Group sizes are taken to be $K_1 = K_2 = 25$ active devices.
The payloads are $w_1 = 128$ and $w_2 = 96$ bits.
The rate of the outer codes are 
$1/2$ and $3/8$ for the first and second class. 
The message from every device is encoded into a block of $n=38400$ channel uses.
The \emph{energy-per-bit} for both groups is identical; and it is denoted by ${E_{\mathrm{b}}}/{N_0}$: this implies that devices in group~1 are using more energy, as they are transmitting more information bits.
The factor graphs for the outer code employed in our simulations are analogous to the designs in~\cite{amalladinne2020unsourced}.
Information messages are partitioned into fragments of 16~bits and, hence, the alphabet size of the outer code in both cases is $2^{16}$.
The factor graphs have parameters $L_1 = L_2 = 16$.
Both $\Am_1$ and $\Am_2$ are formed by sampling rows of Hadamard matrices (excluding rows of all ones).
%

The algorithm performs one round of
BP, after reinitialization, per AMP composite step.
Under the parameters above,  the AMP algorithm converges rapidly.
The disambiguation is subsequently performed on the factor graph of the outer code.
The $K_g$ most promising codewords are returned as output of the last step.
Results do not use phased decoding where top candidates are identified and peeled before repeating the overall CCS-AMP decoding process.
This approach 
 provides  a slight improvement in performance, 
but comes at a substantial cost in terms of computational complexity.
The three implementations described in this section feature essentially equivalent computational complexity.

\begin{figure}[t]
 \centerline{\input{Figures/MultiClassCCS}}
  \caption{This plot shows a comparison of per-user error rate for the two groups under the three candidate implementations.
  The proposed novel multi-class URA algorithm performs best at essentially no additional computation load.}
  \label{fig:errorRatesK100trials}
\end{figure}
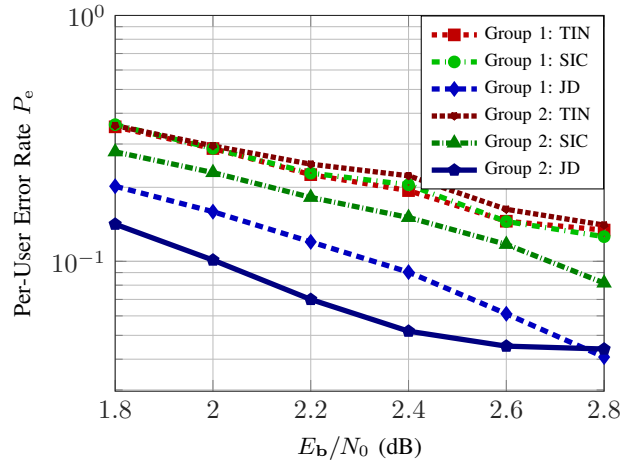
Illustrative numerical results are summarized in Fig.~\ref{fig:errorRatesK100trials}, where per-user probabilities of error are plotted for both groups and the three candidate implementations as a function of energy-per-bit ${E_{\mathrm{b}}}/{N_0}$.
Every point on this graph is averaged over $100$ trials for statistical accuracy.
As expected, group SIC outperforms group TIN;
yet we emphasize that the group~$1$, which is decoded first in SIC, experiences the exact same performance under these two schemes.
Moreover, the multi-class CCS-AMP algorithm performs significantly better than the previous two schemes.
This is hardly surprising because joint decoding typically outperforms sequential decoding schemes or TIN-based approaches.
However, we emphasize two important aspects of our findings.
First, the form of the Onsager in the AMP iterate retains a simple mathematical form despite the presence of the two classes and the coupled state evolution equation.
Second and more importantly, the greater performance of joint decoding comes at essentially no additional complexity.
The same operations must be performed in sequence for all three implementations.
This particular aspect is somewhat surprising because joint decoding schemes are typically much more demanding from a computational point of view.
In closing, it is worth noting that Fig.~\ref{fig:errorRatesK100trials} contains no benchmarks, other than those produced within this work.
This is a byproduct of the fact that standard demixing algorithms are not meant to work at dimensions on the order of $2^{96}$ or $2^{128}$.
%

\section{Conclusion}
In this paper, an extension of coded compressed sensing to the multi-class unsourced random access scenario is studied. 
A joint decoding approach is proposed to concurrently decode messages emanating from multiple classes of users, leveraging insights from compressed demixing.
Within the framework of approximate message passing (AMP), the computational complexity of the proposed scheme is the same as that of alternate implementations such as \textit{treating interference as noise} and \textit{successive interference cancellation}.
However, these other schemes perform poorly compared to concurrent multi-class decoding.
The ability to decode classes of device jointly with essentially no added complexity is surprising and should be embraced.
Our findings provide a blueprint for generic demixing in very high dimensional spaces, which forms a promising avenue for future research.
%
%
%
%
%
%
\newpage

\bibliographystyle{IEEEbib}
\bibliography{IEEEabrv,isit2021}

\end{document}

%% file: preamble.tex
\usepackage{amsmath,amssymb,amsthm}
\usepackage{enumerate}
\usepackage{stfloats}
\usepackage{comment}
\usepackage{subcaption}
\usepackage{siunitx}
\usepackage{mathtools}
\usepackage{accents,latexsym,cancel}
\usepackage{cite}

\usepackage[dvipsnames]{xcolor}
\definecolor{myPink}{RGB}{255,105,183}

\usepackage[T1]{fontenc}
\usepackage{graphics} 
\usepackage[mathscr]{euscript}
\usepackage{algorithm}
\usepackage[noend]{algpseudocode}
\usepackage{bbm}
\makeatletter
\def\BState{\State\hskip-\ALG@thistlm}
\makeatother

\usepackage{tikz}
\usetikzlibrary{arrows,shapes,chains,matrix,positioning,scopes,patterns,calc,
decorations.markings,
decorations.pathmorphing,
}

\usepackage{pgfplots}
\pgfplotsset{compat=1.3}
\usepgflibrary{shapes}

\newtheorem{theorem}{Theorem}

\newtheorem{remark}[theorem]{Remark}

\renewcommand{\epsilon}{\varepsilon}

\newcommand{\RNum}[1]{\uppercase\expandafter{\romannumeral #1\relax}}

\newcommand{\bv}{\ensuremath{\underline{b}}}

\newcommand{\ev}{\ensuremath{\mathbf{e}}}

\newcommand{\kv}{\ensuremath{\mathbf{k}}}

\newcommand{\mv}{\ensuremath{\mathbf{m}}}

\newcommand{\qv}{\ensuremath{\mathbf{q}}}
\newcommand{\rv}{\ensuremath{\mathbf{r}}}
\newcommand{\sv}{\ensuremath{\mathbf{s}}}
\newcommand{\Sv}{\ensuremath{\mathbf{S}}}

\newcommand{\vv}{\ensuremath{\mathbf{v}}}

\newcommand{\wv}{\ensuremath{\mathbf{w}}}
\newcommand{\xv}{\ensuremath{\mathbf{x}}}
\newcommand{\yv}{\ensuremath{\mathbf{y}}}
\newcommand{\zv}{\ensuremath{\mathbf{z}}}

\newcommand{\zerov}{\ensuremath{\mathbf{0}}}

\newcommand{\etav}{\ensuremath{\boldsymbol{\eta}}}

\newcommand{\lambdav}{\ensuremath{\boldsymbol{\lambda}}}

\newcommand{\muv}{\ensuremath{\boldsymbol{\mu}}}

\newcommand{\zetav}{\ensuremath{\boldsymbol{\zeta}}}

\newcommand{\Gv}{\ensuremath{\boldsymbol{\gamma}}}

\newcommand{\Phim}{\ensuremath{\boldsymbol{\Phi}}}
\newcommand{\Am}{\ensuremath{\mathbf{A}}}
\newcommand{\Dm}{\ensuremath{\mathbf{D}}}
\newcommand{\IDm}{\ensuremath{\mathbf{I}}}

\newcommand{\Xv}{\vec{{X}}}

\def \Mc{M_{\mathrm{c}}}

\def \Nc{N_{\mathrm{c}}}

\def \Pc{P_{\mathrm{c}}}

\def \Bc{B_{\mathrm{c}}}

\def\Pr{\mathrm{Pr}}

\DeclareMathAlphabet{\mcl}{OMS}{cmsy}{m}{n}

\newcommand{\wh}{\widehat}

\newlength\tikzwidth
\newlength\tikzheight

\definecolor{mycolor1}{rgb}{0.63529,0.07843,0.18431}%
\definecolor{mycolor2}{rgb}{0.00000,0.44706,0.74118}%
\definecolor{mycolor3}{rgb}{0.00000,0.49804,0.00000}%
\definecolor{mycolor4}{rgb}{0.87059,0.49020,0.00000}%
\definecolor{mycolor5}{rgb}{0.00000,0.44700,0.74100}%
\definecolor{mycolor6}{rgb}{0.74902,0.00000,0.74902}%


%% file: Figures/LDPC.tex
\begin{tikzpicture}
  [
  font=\small, >=stealth', line width=1pt, draw=black,
  check/.style={rectangle, minimum height=4.5mm, minimum width=4.5mm, draw=black, fill=gray!20},
  section/.style={circle, minimum size=5mm, draw=black}
  ]

\node[rotate=90] (message) at (-2.25,0) {Message $\wv$};
\draw[->] (message.east) to [out=90,in=135] (-0.5,1.5);

\foreach \l in {1,2} {
  \node[section] (s\l) at (0.5,2.3-0.9*\l) {};
  \node (v\l) at (1.4,2.3-0.9*\l) {$\vv(\l)$};
}
\node (s4) at (0.5,-0.2) {$\vdots$};
\node (v4) at (1.4,-0.2) {$\vdots$};
\node[section] (sL) at (0.5,-1) {};
\node (vL) at (1.4,-1) {$\vv(L)$};

\node[check] (a1) at (-1.25,0.9) {};
\node[check] (ap) at (-1.25,-0.5) {};
\node (a2) at (-1.25,0.3) {$\vdots$};

\node (variable) at (-0.5,-1.8) {Bipartite Graph};
\node (sparc) at (1.125,-1.8) {$\vv$};
\node (sparc) at (4.3,-1.8) {$\mv$};
\node (index) at (2.5,-1.8) {Indexing};

\draw (s1) -- (a1.east);
\draw (s2) -- (a1.east);
\draw (sL) -- (a1.east);
\draw (s1) -- (ap.east);
\draw (sL) -- (ap.east);

\foreach \l in {1,2} {
  \draw[draw=black] (4.2,2.75-0.9*\l) rectangle (4.4,1.85-0.9*\l);
  \node (m\l) at (3.6,2.3-0.9*\l) {$\mv(\l)$};
}
\node (d4) at (3.6,-0.2) {$\vdots$};
\draw[draw=black] (4.2,3.95-0.9*5) rectangle (4.4,3.05-0.9*5);
\node (mL) at (3.6,-1) {$\mv(L)$};

\draw[->] (v1) to node[above]{$f$} (m1);
\draw[->] (v2) to node[above]{$f$} (m2);
\draw[->] (vL) to node[above]{$f$} (mL);

\end{tikzpicture}

%% file: Figures/BP.tex
\begin{tikzpicture}
  [
  font=\small, line width=1pt, draw=black,
  check/.style={rectangle, minimum height=5mm, minimum width=5mm, draw=black, fill=gray!20},
  trivialcheck/.style={rectangle, minimum height=4mm, minimum width=4mm, draw=black, fill=gray!20},
  section/.style={circle, minimum size=5.5mm, draw=black}
  ]

\foreach \m in {1,2} {
  \node[section] (s\m) at (0,2.7-0.9*\m) {$s_{\m}$};
}
\node (s3) at (0,0.1) {$\vdots$};
\node[section] (s4) at (0,-0.9) {$s_L$};

\foreach \t in {1,2} {
  \node[trivialcheck] (t\t) at (-1.5,2.7-0.9*\t) {}
    edge (s\t);
}
\node (t4) at (-1.5,0.1) {$\vdots$};
\node[trivialcheck] (t5) at (-1.5,-0.9) {}
    edge (s4);

\node[check] (a1) at (2.5,1.35) {$a_1$};
\node (a2) at (2.5,0.55) {$\vdots$};
\node[check] (a3) at (2.5,-0.45) {$a_P$};

\node[rotate=90] (variable) at (-2.2,0.45) {Local Observations};
\node[rotate=90] (check) at (3.7,0.45) {FFT-based Factors};

\draw (s1) -- (a1.west);
\draw (s2) -- (a1.west);
\draw (s4) -- (a1.west);
\draw (s1) -- (a3.west);
\draw (s4) -- (a3.west);

\draw[shorten <=0.8cm,shorten >=0.65cm,->] (s1)++(0,0.2cm) -- node[above,rotate=-12.5] {$\muv_{s_1 \to a_1}$} ([yshift=0.2cm]a1.west);
\draw[shorten <=0.7cm,shorten >=0.75cm,<-] (s4)++(0,-0.2cm) -- node[below,rotate=12.5] {$\muv_{a_P \to s_L}$} ([yshift=-0.2cm]a3.west);
\draw[shorten <=0.4cm,shorten >=0.4cm,->] (t2)++(0,0.2cm) -- node[above] {$\lambdav_2$} (-0.2,1.1cm);

\end{tikzpicture}

%% file: Figures/MultiClassCCS.tex
        
\begin{tikzpicture}
\definecolor{colorTIN2}{RGB}{128,0,0}
\definecolor{colorTIN1}{RGB}{192,0,0}
\definecolor{colorSIC2}{RGB}{0,128,0}
\definecolor{colorSIC1}{RGB}{0,192,0}
\definecolor{colorJD2}{RGB}{0,0,128}
\definecolor{colorJD1}{RGB}{0,0,192}

\begin{semilogyaxis}[
font=\small,
width=6.5cm,
height=5cm,
scale only axis,
every outer x axis line/.append style={white!15!black},
every x tick label/.append style={font=\color{white!15!black}},
xmin=1.8,
xmax=2.8,
xtick={1.8,2,2.2,2.4,...,2.8},
xlabel={$E_{\mathbf{b}}/N_0$ (dB)},
xmajorgrids,
xminorgrids,
every outer y axis line/.append style={white!15!black},
every y tick label/.append style={font=\color{white!15!black}},
ymin=0,
ymax=1,
ytick={0.01, 0.1, 1.0},
ylabel={Per-User Error Rate $P_{\mathrm{e}}$},
ymajorgrids,
yminorgrids,
legend style={at={(1, 1)},anchor=north east,draw=black, fill=white, legend cell align=left,font=\scriptsize}
]

\addplot [color=colorTIN1, dotted,line width=2.0pt,mark size=1.4pt,mark=square,mark options={solid}]
  table[row sep=crcr]{
  1.8 0.3524\\
  2 0.2876\\
  2.2 0.2252\\
  2.4 0.194\\
  2.6 0.1456\\
  2.8 0.134\\
};
\addlegendentry{Group 1: TIN};

\addplot [color=colorSIC1,dashdotted,line width=2.0pt,mark size=1.4pt,mark=o,mark options={solid}]
  table[row sep=crcr]{
  1.8 0.3596\\
  2 0.288\\
  2.2 0.2284\\
  2.4 0.2048\\
  2.6 0.1452\\
  2.8 0.126\\ 
};
\addlegendentry{Group 1: SIC};

\addplot [color=colorJD1,densely dashed,line width=2.0pt,mark size=1.4pt,mark=diamond,mark options={solid}]
  table[row sep=crcr]{
  1.8 0.202\\
  2 0.1592\\
  2.2 0.12\\
  2.4 0.0904\\
  2.6 0.0612\\
  2.8 0.0408\\
};
\addlegendentry{Group 1: JD};

\addplot [color=colorTIN2,densely dotted,line width=2.0pt,mark size=1.4pt,mark=star,mark options={solid}]
  table[row sep=crcr]{
  1.8 0.3564\\
  2 0.2948\\
  2.2 0.2484\\
  2.4 0.2228\\
  2.6 0.1628\\
  2.8 0.1404\\
};
\addlegendentry{Group 2: TIN};

\addplot [color=colorSIC2,densely dashdotted,line width=2.0pt,mark size=1.4pt,mark=triangle,mark options={solid}]
  table[row sep=crcr]{
  1.8 0.2788\\
  2 0.2296\\
  2.2 0.182\\
  2.4 0.1508\\
  2.6 0.1172\\
  2.8 0.0816\\
};
\addlegendentry{Group 2: SIC};

\addplot [color=colorJD2,solid,line width=2.0pt,mark size=1.4pt,mark=pentagon,mark options={solid}]
  table[row sep=crcr]{
  1.8 0.1416\\
  2 0.1012\\
  2.2 0.07\\
  2.4 0.052\\
  2.6 0.0452\\
  2.8 0.044\\
};
\addlegendentry{Group 2: JD};

\end{semilogyaxis}
\end{tikzpicture}